\documentclass{aa}
\usepackage{graphicx}
\usepackage{natbib}
\usepackage{txfonts}

\def\new#1 {{\bf #1 }}
\def\cut#1 {\sout{#1} }

\def\vt{V773 Tau~A~}
\def\beq{\begin{equation}}
\def\eneq{\end{equation}}
\def\simgt{\lower.5ex\hbox{$\; \buildrel > \over \sim \;$}}
\def\simlt{\lower.5ex\hbox{$\; \buildrel < \over \sim \;$}}

\begin{document}

\title{
Synchrotron  emission  from  the T Tauri binary system V773 Tau A}

\author{M. Massi   \inst{1},  J. Forbrich   \inst{1},  K. M. Menten   \inst{1},
G. Torricelli-Ciamponi   \inst{2}, J. Neidh{\"o}fer   \inst{1}, S. Leurini   \inst{1}, and
F. Bertoldi   \inst{3}}

\institute{Max-Planck-Institut f\"ur Radioastronomie,
Auf dem H\"ugel 69, D-53121 Bonn, Germany\\
\and INAF, Osservatorio Astrofisico di Arcetri, Largo E. Fermi 5,
I-50125 Firenze, Italy\\
\and Radioastronomisches Institut, Universit\"at Bonn,
Auf dem H\"ugel 71, Bonn, D-53121, Germany
}

\offprints{M. Massi, \\ \email{mmassi@mpifr-bonn.mpg.de}}

\date{Accepted 06/03/2006 }

\abstract{
The  pre-main sequence binary system \object{V773 Tau A} shows  remarkable
flaring activity around  periastron passage.
Here, we present the  observation of such  a flare at a wavelength of 3 mm
(90 GHz)  performed with the
Plateau de Bure Interferometer.%
%
\thanks
{Based on observations carried out with the IRAM Plateau de Bure Interferometer. IRAM is supported by INSU/CNRS (France), MPG (Germany), and IGN (Spain).}
We examine different possible causes for the energy losses responsible for
the  e-folding time of 2.31$ \pm$ 0.19 hours of that flare.
We exclude synchrotron, collisional, and inverse Compton losses
because they are not consistent with observational constraints,
and we propose that the fading of the emission is due to the leakage of
electrons themselves  at each reflection between the two mirror points
of the  magnetic structure partially trapping them.
The magnetic structure compatible with both our leakage model   and
previous observations  is that of a helmet streamer that, as in the solar
case, can occur
at the top of the X-ray-emitting, stellar-sized coronal loops of one of the stars.
The streamer may extend up to 
$\sim$ 20 $R_*$  and  interact with the corona of the other star at
periastron passage, causing  recurring flares.
The inferred  magnetic field strength at the two mirror points
of the helmet streamer is in the range 0.12 - 125 G,
and the corresponding  Lorentz factor, $\gamma$, of the partially trapped  electrons
is in the range 20 $< \gamma <$ 632. We therefore
rule out that the emission could be of
gyro-synchrotron nature: the derived high  Lorentz factor
proves  that the nature of the emission at 90 GHz from this pre-main binary system
is  synchrotron radiation.
}

\titlerunning{
Synchrotron  emission  from  a T Tauri binary system
}

\authorrunning{Massi et al.}

\maketitle

\section{Introduction} \label{introduction}

The dynamo
theory  (Parker 1955) explains how differential rotation  generates
a toroidal field in
the  interior of a star from an initial stellar  dipole field, and how
convection,  bringing  this field  up to the surface,
creates the coronal magnetic arc-like structures called loops.
A close relationship between
magnetic loops and  flares exists.  In fact,
the physical mechanism 
invoked to explain solar and stellar flares
today is  magnetic reconnection,
which occurs  when  different loops interact with each other
or when  field lines of the same  loop are stretched.
The released  magnetic energy   accelerates
a part of the thermal electrons trapped in the loops
to high energies, causing the flare
(Golub \& Pasachoff 1997; Priest \& Forbes 2002).

We have observed  a strong flare  at mm
wavelengths toward
the T Tauri binary system V773 Tau A. In the context of
the  outlined scenario, our aim is to investigate: {\it a}) the flare
location and hence
the geometrical structure of the involved magnetic field, and
{\it b}) the nature of the observed emission, i.e. whether  the
magnetic field intensity and the electron Lorentz factor can be 
derived, yielding information on the emission mechanism.

Concerning  the first issue, {\it a},  one
can distinguish among
three  possible origins for flares:
\begin{enumerate}
\item
In a \textit{single} star, flares
 occur when
new  emerging loops from below the stellar surface
move
into older, already-existing loops (Heyvaerts et al. 1977; Massi et al. 2005).
Such intruding loops have indeed been observed on the Sun in 
 high-resolution maps obtained
with the Nobeyama Radio Interferometer (Nishio et al. 1996).
\item
In  protostars or young (pre-main sequence) stellar systems,
which  are still surrounded by a dense accretion disk,
flares can occur because of
stretching, disruption, and reconnection of
magnetic field lines between the star and its disk
(Feigelson \&\ Montmerle 1999).
\item
In  close binary systems
consisting of  late spectral type stars,
much higher  magnetic activity  is expected
compared to the Sun (because tidal synchronism
increases  the
rotation rate and therefore the dynamo efficiency). Moreover,
the coronal loops of the two stars can interact with each other,
as for RS CVn (close) binaries
(Uchida \& Sakurai 1983; Graffagnino et al. 1995).
Observational evidence  for   inter-binary loop ``collisions"
  was  also recently found  for the  young binary  system V773 Tau A,
which shows a high rate of radio  flares around the periastron passage
(Massi et al. 2002).
\end{enumerate}

Concerning the second issue ({\it b}),
the usually observed  radio emission from stellar coronae
shows spectra peaking around
10 GHz and is consistent with the gyro-synchrotron process
(Mutel et al. 1985; Bastian et al. 1998; G\"udel 2002).
However,  some solar and stellar flares  can also be observed at
millimeter wavelengths (Fig. 1), and the
spectra are thought  to be due to the   superposition of a
gyro-synchrotron spectrum  with a synchrotron spectrum,
the last one peaking in the shorter submillimeter to the far-infrared range
(Kaufmann et al. 1988; Kaufmann et al. 2004).
Gyro-synchrotron emission, associated with  mildly
relativistic particles,
i.e., with a Lorentz factor of $\gamma < 10$,
is  circularly   polarized.
In contrast, synchrotron emission comes from  highly relativistic particles ($\gamma \gg 1$)
and is linearly polarized
(Dulk 1985; Phillips et al. 1996;  Tsuboi et al. 1998).

To probe the existence of relativistic electrons
in the \vt system and to
constrain their energy and the strength and  topology
of the magnetic field, we observed it
at millimeter wavelengths using the IRAM Plateau de Bure
Interferometer.

The paper is organized as follows.
\S \ref{source} reviews previous results on
\vt whereas
in \S \ref{obs} our new
observations and results are presented.
In \S  \ref{losses} we evaluate which kind of  energetic
processes may
cause the trapped electrons
to lose their energy  and  derive
the Lorentz factor and magnetic field strength. These deductions
are confirmed in \S \ref{rising} by the analysis of the flare rising
time.
\S \ref{loop}  discusses the  magnetic field configuration,
while  \S \ref{conclusions}  presents our conclusions.
\begin{figure}[htb]
\centering
\includegraphics[height=8cm, angle=-90]{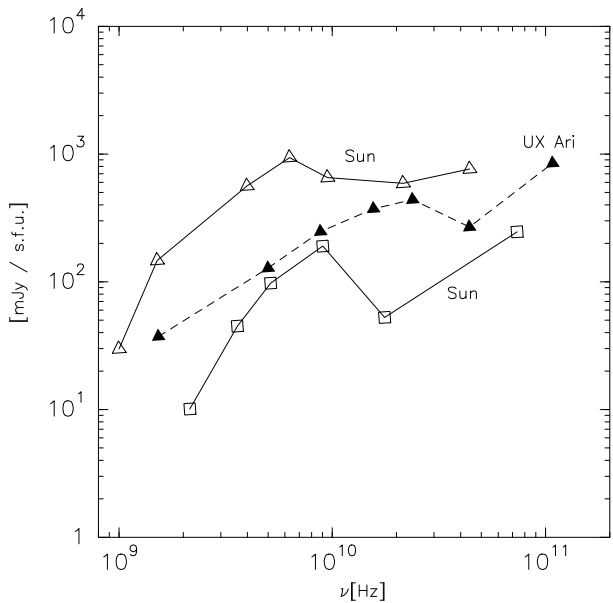}
\caption{
Examples of spectra of the Sun
(squares: Akabane et al. 1973; empty triangles: Zirin and Tanaka 1973)
and UX Arietis (filled triangles: Beasley \& Bastian 1998)
exhibiting  a ``flattening'' towards mm-wavelengths.
Notice that the vertical scale is in both solar flux units (for the solar
spectra) and mJy (for UX Arietis).
}
\label{kauf}
\end{figure}

\section{The \vt  binary system} \label{source}
V773 Tau,
at a  distance  of 148$\pm$5 pc (Lestrade et al. 1999),
is a quadruplet of T Tauri stars within an area  of radius
less than 100 AU (Woitas 2003).
The system has  two   stars  with  orbital periods of years
and two shorter period inner stars.  The object of our studies
is the V773 Tau A binary, with
an orbital period of only  51.075 days (Welty 1995).
The orbit of V773 Tau A is moderately eccentric (e=0.3)
and the epoch of periastron passage, corresponding to orbital
phase $\Phi=0$, is $t_0$=2449330.94 JD.
The stars, both very  active,
have similar radii ($R_*$= 2.4 $R_{\odot}$) and
rotation periods (about 3 days), but seem to have different masses
(see Welty 1995).

The system possesses only weak hydrogen emission lines in its spectrum,
which implies that its components are weak-line T Tauri stars 
with mass loss. This follows from the presence of forbidden emission lines
(Cabrit et al. 1990).
The presence of a disk is controversial (Skinner et al. 1997) since the
observed infrared excess may  also be attributable to V773 Tau D,
the forth member of the quadruplet (Duch\^ene et al. 2003).
Also, it is difficult to establish any thermal component
in
the millimeter emission (which would imply a disk),
since it  is  strongly dominated by a variable non-thermal component
typical of flaring coronal activity
(Skinner et al. 1997).
At 2.7 mm,
variability was found
with the flux density changing  from 30 mJy to less than 3 mJy
in a period of a few months (see  Skinner et al. 1997).
Beckwith et al. (1990)
measured a flux density at 1.3 mm, $S = 42 \pm 6$ mJy,
whereas an observation several years later gave
$S = 24 \pm 4$ mJy (see  Skinner et al. 1997).
With the Effelsberg 100-m  telescope,
Massi, Menten, \&  Neidh\"ofer (2002)
observed  a flare at 7 mm
with  $S = 68\pm 13$ mJy,
and three months later they only obtained
 an upper limit (3$\sigma$)  of
23 mJy.

\vt is not the only pre-main sequence object with mm-flaring
emission.
For example, Bower et al. (2003) observed a giant outburst at 86 GHz
in the Orion source GMR-A, also a weak-line T Tauri star.
Long-lasting (13 days) millimeter  activity
in that source was observed by Furuya et al.  (2003).
However, V773 Tau A is the only known system with coronal solar-like magnetic
activity where linearly polarized emission
together  with the more common
circularly  polarized emission was measured (Phillips et al. 1996).
The high level of variable millimeter emission together with
the presence of linear  polarization
indicates that
a quite  energetic electron population is present.
The fact that
these energetic electrons have  rather short lifetimes is
worthy of note for the following analysis.
Phillips et al. (1996) find that
total intensity variations  at different wavelengths and epochs
share common time scales and
establish a recurring interval for the  decay
of radio events with
a typical    1/e   time scale of only about 1 hour.

The two components of \vt are both active
(Welty 1995);
flares on both of them are expected due to  mechanism
$\it {a}$-1 discussed in the introduction. In addition, however,
there is evidence for mechanism $\it {a}$-3 (inter-binary loop collision):
by folding the data, monitored over 522 days,
with the  orbital period
(51.075 days), the flares appear to be clustered at  periastron passage
(Massi et al. 2002).
Around  the periastron,   more than one interbinary
interaction seems to occur:
three consecutive flares, separated by a time interval of
3 -- 4 days, have been observed during a continuous observation
around periastron passage
(Fig. 3 in Massi et al.  2002).
The hypothesis  of a stable  active region
simply eclipsed by the body of the star during each rotation ($P_{\rm {rot}}$= 3 days)
is ruled out by the short lifetimes (1 -- 2  hour only)
of the energetic electrons
(as discussed above and in \S 3).
On the contrary, a large asymmetric  structure rooted on one of
the two  rotating stars
could  explain the occurrence of
the three consecutive flares,
as repeated  collisions (at each rotation) 
with the corona of the other star.
The extent, $H$, of this  magnetic structure must be large
enough to allow the observed consecutive  collisions  around
periastron passage.
The periastron separation is 56 solar radii, $R_{\odot}$,
which for  a stellar radius, $R_*=2.4~R_{\odot}$ (Welty 1995),
corresponds to  23 $R_*$.
Indeed, a large magnetic structure has been imaged by
using Very Long Baseline Interferometry  (VLBI, Phillips et al. 1996; orbital
phase $\phi=0.1$)
at
$\lambda=3.6$ cm, with
two peaks  of different intensity separated by 0.17 AU (15 $R_*$).
Observational constraints for the size of the emitting region
are therefore in the range 15~--~20~$R_*$.

The fact that
linear polarization has been observed
implies that   Faraday depolarization is not effective and
puts
an upper limit to the  density ($n < 10^9$ cm$^{-3}$)
of the  plasma confined in  this  extended magnetic structure
(Phillips et al. 1996).
X-ray observations give  evidence not only of
a density much above this limit, but also
of the existence of smaller,
solar-like, coronal loops in V773 Tau A.
Skinner et al. (1997)
interpreted
the light curve of a hard X-ray flare in \vt
as being due to the rotational modulation of the emitting flaring region,
determining  a size of $H\le 0.6\, R_*$.
Tsuboi et al. (1998) interpreted the decay of another hard X-ray flare
as being due to radiative cooling,   obtaining  a  size
of $1.4\, R_*$.
Therefore,   radio and X-ray emission come from spatially separated regions:
denser and smaller
ones associated with X-ray emission and 
larger and more diffuse ones associated with radio emission.
Further confirmation of the presence of two structures comes from
the multiwavelength campaign on V773 Tau A
carried out by Feigelson et al. (1994)
showing    radio variability combined with
a  steady X-ray flux. 
Using our millimeter observations,
the following sections are aimed to better define
what kind of extended magnetic structure  might confine
the radio-emitting  plasma, the intensity of the magnetic field,
and  the energy of the trapped particles.

\section{Observations and data reduction} \label{obs}

\vt was observed with the IRAM Plateau de Bure Interferometer
at 1 mm and 3 mm
around the two  periastron passages in  August
and November 2003.
The configuration was 5D (five antennas) in August
and 6Cp (the six-antenna configuration)
in November.
Observations
were done in LCP for the 3 mm observations and in RCP for the 1 mm observations.
The continuum was measured with four correlator units per
receiver, twice with a bandwidth of 320~MHz and twice with a bandwidth of
160~MHz.
Cyclical observations (every 20 minutes) of the sources
0528+134 and 3C286 were used for phase and amplitude calibrations.
The data were reduced with the programs  CLIC and MAPPING
of the GILDAS software package developed by the Grenoble
Astrophysics Group. The source was always unresolved.
A complete flare (Fig. 2) was observed on August 6 ($\Phi=0.1$).
With about 12 mm of precipitable water vapor in the
atmosphere, no  observations at
$\lambda = 1$~mm were possible, and those at $\lambda = 3$~mm
had  the high rms noise level of 17 mJy; the 3 mm receiver was tuned to
90 GHz and had a system temperature   around 1000~K.
The best power-law fit to the decay part of the light curve
shown in Fig. 2 gives  an e-folding time of $\tau = 2.31 \pm 0.19$ hour.
Around the   periastron passage in November,
we were able to observe under  very good weather conditions
for two consecutive days at $\Phi=0$.
The flux density was at a quiescent constant level
with a nearly  flat spectrum (Table 1).

\begin{figure}[h]
\centering
\resizebox{\hsize}{!}{\includegraphics[ angle=-90]{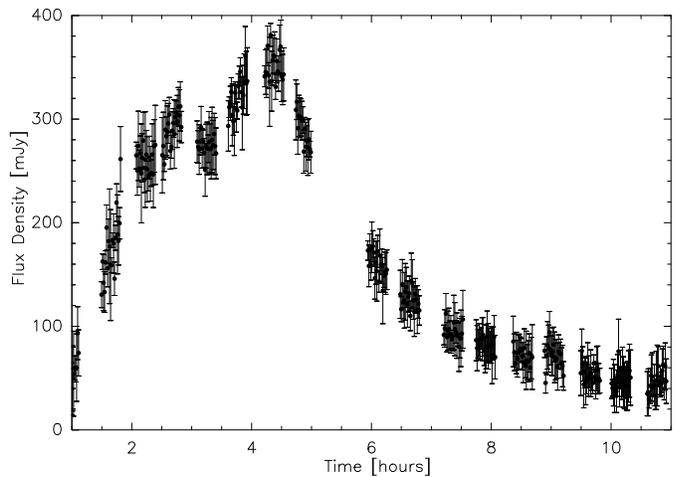}}
\caption{
V773 Tau A
flare  observed  August 6, 2003 ($\Phi=0.1$) with the Plateau de Bure Interferometer
at $\lambda = 3$~mm.
The error bar corresponds to 2$\sigma$.
}
\label{}
\end{figure}
\begin{table}
\begin{center}
\caption[]{Flux density in   November 2003}
\begin{tabular}{lcc}
\hline \hline \noalign{\smallskip}
Day& $S_{\rm 1 mm}$ (mJy) & $S_{\rm 3 mm}$ (mJy)  \\
\noalign{\smallskip} \hline \noalign{\smallskip}
13& 5.0 $\pm$ 1.2 &3.0  $\pm$ 0.4      \\
14& 6.0 $\pm$ 1.2 &5.0  $\pm$ 0.3      \\
\noalign{\smallskip} \hline
\end{tabular}
\end{center}
\label{table:pdm}
\end{table}

\section{Rapid flare decay: leakage of the emitting particles} \label{losses}

As outlined in the introduction, the emission observed at 3.3~mm (90 GHz)
can be interpreted as synchrotron radiation. In synchrotron radiation, the emission of each
relativistic  electron with Lorentz factor $\gamma$
moving in a magnetic field $B$ (in Gauss)  is centered
around  a peak spectral
frequency, $\nu_{\rm 0}$
(Ginzburg and Syrovatski 1965) of
\beq
\nu_{\rm 0}= 1.8 \times 10^{6} B \gamma ^2 ~~~{\rm Hz}.
\label{peak}
\eneq
This  implies that,  to reproduce the observed emission at
$\nu = 90$ GHz,  electrons must exist for which
the following relationship holds:
\beq
B \gamma ^2=5 \times 10^4.
\label{b_ga}
\eneq

The temporal evolution of the flux density shown in Fig. 2 clearly
indicates that
the emitting electron distribution, responsible for the observed
synchrotron emission, is subject to  some  losses. In particular,
the derived e-folding time of 2.31 hours for the mm flare is of the same order
of that 
measured by Phillips et al. (1996)  for cm emission (about one  hour).
This time scale  implies that the  electrons
must be  at least partially  trapped
in the emitting region, which would  otherwise be
rapidly  depleted in a time equal to the size of the region divided
by the speed of the electrons:
$20 R_*/c \sim  112$ s.
It is well-established that particles trapped in magnetic
structures  spiral around field lines to a point (the ``mirror point'')
where the magnetic field lines converge and the
field strength is sufficient to cause the particles  to reverse direction
and travel back to the other mirror point.
Reflected  back and forth
between mirror points (known as ``bounce'' motion), the particles
remain trapped  and, continuously spiraling, emit synchrotron radiation
until they have lost their energy (Roederer 1970).

However,
the decay time of 2.31 hours is not attributable to synchrotron losses.
As derived in the  Appendix,
synchrotron losses would explain an intensity
decay time of that duration only for
an emitting region of size  $\le$ 3.8 R$_*$, whereas  the  observed
size  (\S 2)  is  $H \ge 15 R{_*}$.
Neither inverse Compton losses, due to the radiation
field of the star, and/or to the synchrotron emission of the flare
(synchrotron self-Compton (SSC)), can be responsible for the observed decay;
Compton losses are even lower than synchrotron losses
(see Appendix). On the other hand, even collisional losses cannot
account for the observed decay time, since
collisions are important
only for high density plasmas (see Appendix) in contrast to
the upper limit on the plasma density $n < 10^9$ cm$^{-3}$
derived from Faraday depolarization (\S 2).

Since the fast decay of the emission cannot be attributed to energetic
losses of the electrons, it may be caused by
leakage of the particles themselves, meaning that at each
bounce motion at one mirror point, some of the electrons
are able to leave the trapping region
and escape into free  space.
This is possible for electrons having
a  pitch angle, $\theta$ (the angle between  their velocity direction
and the magnetic field vector),  smaller than the loss cone angle
\beq
{\rm sin}\,\theta< {\rm sin}\,\theta_0=  \left ({B_1 \over B_0}\right)^{1/2}.
\eneq
Here, $B_1$ is the magnetic field intensity  far out in the
corona at a distance $(H_1)$, while $B_0$ is that at a lower distance
$(H_0)$.
Pitch angle scattering, also known as ``pitch angle diffusion"
(Melrose \& Brown 1976),
is one of the main processes by which  trapped particles are lost.
Weak and strong diffusion have been introduced by Melrose \& Brown
(1976).
However, in order to be effective, weak diffusion demands
a high-density medium
($n>10^9 {\rm cm} ^{-3}$), in contrast  to what we discussed above.
Also, Lee et al. (2002) found that
for the microwave decay of a solar flare, the
assumption of strong diffusion seems to be the
most appropriate.
In this scenario, the emission decreases by a factor of 1/e if
the precipitation rate
\beq
\nu_{\rm p}= {1 \over 2} \theta_0^2  {c\over (H_1-H_0)}
\eneq
(Melrose \& Brown 1976; Lee et al. 2002)
times the decay time (2.31 hours, in our case) equals unity.
This condition, valid for small $\theta_0$ values, implies:
\beq
742 \times {B_1\over B_0}~{R_*\over  H_1-H_0}=1.
\label{trap}
\eneq
For a magnetic dipole field, ${B_1/ B_0}=[{H_0 / H_1}]^{3}$,
possible solutions of Eq. (\ref{trap}) are  values in the
ranges $H_0 \sim (2-5) R_*$ and $H_1\sim (10-20) R_*$.
Assuming  a magnetic field strength at the stellar surface
of the order of 1 kG (see Appendix), 
the  corresponding values of the magnetic
field intensity   are in the range $125 - 8$ G  for $H_0$ and
  $1 - 0.12$ G for $H_1$.
This scenario is consistent with synchrotron emission,
since from Eq. (\ref{b_ga}),  the corresponding
Lorentz factor of
relativistic electrons emitting at 90 GHz is   20$< \gamma <$ 632.

This scenario  is also in accordance
with the VLBI observation of Phillips et al. (1996),
since the analogon of Eq. (\ref{trap}) derived for the decay time
of one hour and solved
for   $H_1-H_0=15R_*$
with  $H_1 \sim 20 R_*$
gives $B$
in the range of $4.6- 0.12$ G. These values of the magnetic field
for  $\nu =8.2$ GHz (Phillips et al. 1996) correspond (see Eq. (1))
to  31 $< \gamma <$ 195
indicating, therefore, the presence  of relativistic  electrons
in agreement with  the observed linear polarization.

\section{The flaring phase: a propagating shock} \label{rising}

In the previous section, we  established the most likely physical process
responsible for the rapid flare decay.
The aim of this section is to test whether the
additional constraint set by the flare rising time  confirms the previously obtained
results.

Recent solar flare studies (Tanuma \& Shibata 2005; Asai et al. 2004)
have shown  that the observed downflow motions
can create fast shocks that are related to  observed non-thermal bursts,
since  shocks are supposed to be sites of electron acceleration (Aschwanden 2002).
In our scenario, in which reconnection takes place far out
where the two  stars' magnetospheres interact (i.e. at $H_1$), the shock induced
by magnetic reconnection  can propagate  down  to $H_0$ along  the magnetic structure
at the local Alfv\'en velocity
\beq
v_A(H)={B(H) \over \sqrt{4 \pi m_H n(H)}}
\simeq 7~10^9
\left ({H \over R_*}\right)^{-2} ~~{\rm cm/sec}
\eneq
(for a dipole magnetic field  with $B(R_*)=1000 $ G and a
particle density $n \sim 10^9 (H/R_*)^{-2}$ cm$^{-3}$),  thus inducing
successive particle acceleration events.
The shock propagation speed can be related to the flare rise time,
$t_{\rm rise}$, 
\beq
t_{\rm rise}= \int^{H_1} _{H_0}{ dH \over v_A(H)},$$
\eneq
from which the following relationship can be derived:
\beq
\left ({H_1 \over R_*}\right)^3= 452 
\left (t_{\rm rise} \over {\rm hour} \right) 
\left ({B(R_*)\over 10^3 \rm{G}} \right) 
\left ({n(R_*)\over 10^9 \rm{cm}^{-3}} \right)^{-{1 \over 2}}
+~ \left ({H_0 \over R_*}\right)^3,
\eneq
For  $t_{\rm rise}, \simeq 4.5$ hours and  $H_0$ between  $(2 - 5)~R_*$,
the resulting  height 
$H_1$ is  in the interval 
of $(13 - 27)~ R_*$, assuming that  $B(R_*)$ is in the range 1000-3000 G (see
Appendix) and the density is in the range $~10^8-10^9$ cm$^{-3}$, as in Fig. 3.
\begin{figure}[h]
\centering
\resizebox{\hsize}{!}{\includegraphics[ angle=0]{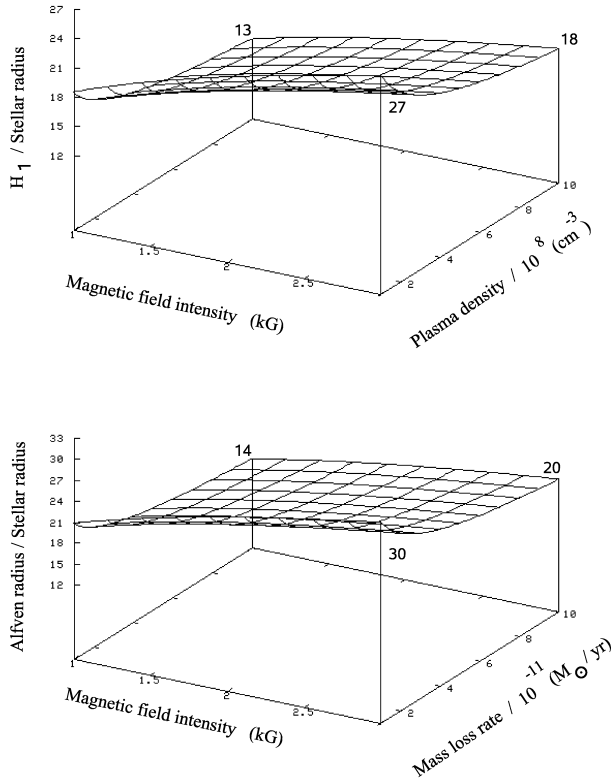}}
\caption{
Top: Variations of the   height H$_1$  (Eq. (8)),
where magnetic reconnection occurs, as a function of magnetic
field intensity and plasma density (see \S \ref{rising}).
Bottom: Alfv\'en radius (Eq. (9)) as a function of magnetic field
intensity and mass loss rate (see \S \ref{loop})}.
\end{figure}
This  range for $H_1$ is 
well-consistent with  the  scenario
of a reconnection event at a 
large  stellar distance and in agreement  with   the resulting
interval   $(10-20) R_*$ from the leakage model.
\section{Magnetic configuration} \label{loop}

In the previous sections, in the context of our leakage model,
we established
magnetic field values  in the range 125 -- 8  G (for  $H_0=2$ -- $5~R_*$) and
0.12 G (at $H_1 = 20~R_*$)
at  the two mirror points of the confining magnetic structure.
Also, we determined values of
4.6 G (at $H_0=6~R_*$) and  0.12 G (at $H_1=20~R_*$)
for the centimeter observations of Philipps et al. (1996).
In both cases,
in a high-resolution image of the source brightness
distribution,
the two mirror points
would correspond to two peaks (displaced by $H_1-H_0$)
of rather different intensity,
because of the relationship between
emissivity and  $B$
(Dulk 1985).
This predicted morphology finds its confirmation
in  the Very Long Baseline Array (VLBA) image
of Phillips et al. (1996).
The VLBA image,
given in the small box of Fig. 3,
shows
a strong  brightness distribution peak, which we may identify with the first
mirror point, $P_0$,
separated by 15 R$_*$ from a second  weaker  peak, which may be
coincident with the second mirror point, $P_1$.

Let us assume the scenario of two stellar coronae interacting at  periastron passage.
One of the two coronae must be very asymmetric,
in accordance with the asymmetric magnetic field configuration
implied by the Massi et al. (2002)
observations
of three consecutive flares around periastron (see  Sect. 2).
On the other hand, a single  giant stellar loop is ruled out by X-ray observations
which, on the contrary (see Sect. 2),
provide evidence for a stellar-sized coronal loop
(Skinner et al. 1997; Tsuboi et al. 1998).
However, in the case of the Sun,
extended magnetic structures are sometimes present above coronal loops.
Indeed, when the Sun is  observed with white-light coronographs, one can
see that
above the top of some coronal loops (Fig. 4),
a dome-shaped structure (the ``helmet") is formed,
extending out to
2 -- 4 solar radii,  with
streamers that can  extend out to many solar radii
(Suess \& Nerney 2004; Endeve et al. 2004).
A helmet streamer is therefore  a plausible physical model for the
elongated asymmetric structure implied by the previous  observations of Massi
et al. (2002).
Moreover, a helmet streamer structure
with its two mirror points at P$_0$ and P$_1$ (Fig. 4) matches
the results  of our leakage model based on  PdBI data, as well as 
the appearence of the source in the VLBA image.

Finally, we point out that the helmet streamer in the simplified
sketch of Fig. 4 is represented, for simplicity's sake, by a straight
structure. In real highly conducting stellar
plasmas, where the ionized gas is frozen
to the field lines, more complex morphologies will occur:
wherever magnetic pressure exceeds gas
pressure, the fluid moves only along the field lines. In contrast,
when  gas pressure dominates,
the  magnetic field lines do move or bend, depending on  the fluid motion.
In a rotating star 
with  mass loss rate $\dot{M}$ and
terminal wind velocity v$_{\infty}$,  the distance
(called Alfv\'en radius, $R_A$)
where  the
corotation of the wind can no longer be enforced by the magnetic field
and the magnetic field lines become curved
is, as given by Andr\'e et al. (1988),
\beq
{R_A \over R_*}= 26 \left ({B_*\over 10^4 \rm{G}} \right) ^{{1 \over 3}}
\left ( {\dot{M} \over 10^{-10} \rm{M}_\odot ~
\rm{yr}^{-1} } \right )^{-{ 1 \over 6}} \left ( {{\rm v}_{\infty} \over 10^3 ~ \rm{km ~ s}
^{-1}} \right )^{{1 \over 6}} \left ({P \over 1 \rm{d}} \right)^{{ 1 \over 3}}.
\eneq
Adopting  a  
range $10^{-11}$--$10^{-10} \rm{M}_{\odot}~  \rm{yr}^{-1}$
for the  mass loss rate of a weak T Tauri star 
(Andr\'e et al.  1992), 
and   a wind terminal velocity of 300 km/sec (Ardila et al.  2002), we can conclude that
the Alfv\'en radius for $P=3$ d  (Fig. 3 bottom) is  in the range 
of $(14 - 30)~R_*$.  

\begin{figure}[h]
\centering
\resizebox{\hsize}{!}{\includegraphics[ angle=0]{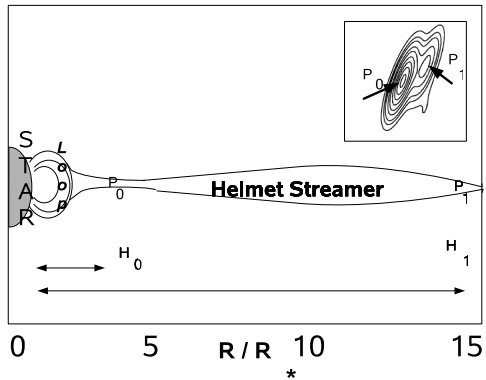}}
\caption{
Sketch of the magnetic structure of a helmet streamer, above a coronal
loop,  as treated in
magnetohydrodynamic models (Endeve et al. 2004).
The small box contains the VLBA image by Phillips et al. (1996).
}
\label{}
\end{figure}

\section{Conclusions} \label{conclusions}

\begin{figure}[h]
\centering
\resizebox{\hsize}{!}{\includegraphics[ angle=-90]{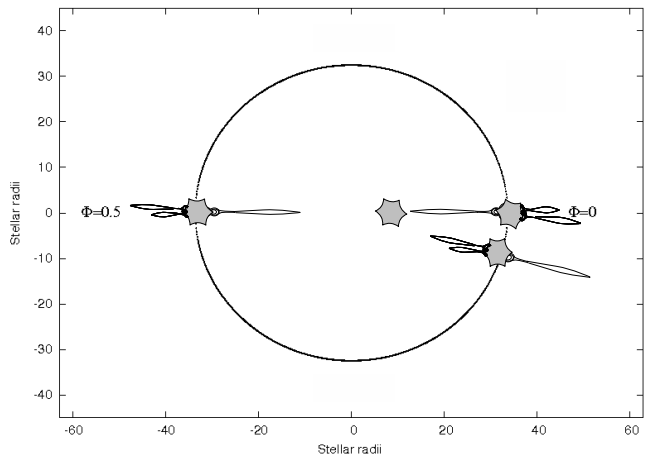}}
\caption{
Sketch of the binary system with
the  orbit parameters of
Welty (1995).
One star with helmet streamers  is shown
at   different positions
along the orbit.
Interbinary collisions  may occur at periastron passage (see text).}
\label{orb}
\end{figure}

%
%
%
Our previous  observations of the binary system 
V773 Tau A (Massi et al.  2002)
have shown inter-binary loop collisions at periastron passage. 
We performed millimeter wavelength observations 
aimed to better define  the
extended magnetic structure   confining the
radio-emitting  plasma,  its  topology, strength,
and  the energy of particles trapped in it.
Based on our results, we propose the following scenario.

Radio and X-ray emission from \vt arise from spatially separated volumes:
X-ray emission arises from denser and smaller regions (size $ \simlt 1 R_*$)
like the closed loop region under P$_0$ in the sketch of Fig. 4
(see \S 2). The radio emission emerges from
larger and   more diffuse regions  that are
similar to the solar helmet streamers.
A helmet streamer from one of the stars,  formed
at the top of a coronal loop, as in the case of the Sun, could  interact with the
corona of the other star and could produce 
the strong radio flares  observed around
periastron passage 
(Massi et al. 2002).
In this scenario, the lack of flares at some periastron passage (as in the 
PdBI observations of  Table 1) could be explained
as the absence of a collision, possibly because
no large enough helmet streamer was emerging on the  hemisphere facing the
other star.
As shown in Fig. 5, because of
the slightly eccentric orbit ($e\simeq 0.3$, Welty 1995)
the distance at periastron
(23 $R_*$)
is  appreciably less than at apoastron (40 $R_*$).
The  size of the helmet streamer should therefore be comparable 
with the periastron distance, but still be less than 40 $R_*$.
In this paper, where we  report  a flare at 3 mm
with $S_{{\rm peak}} = 360~\pm 17~$mJy, we   
have derived 
the size ``$H_1$"
of the emitting region (i.e. the helmet streamer)
in three completely independent ways. 
Interpreting  
the  rising time of the flare in terms of propagation towards the low corona
of a shock due to  a magnetic    
reconnection event occurring at 
large height, we  have determined a  height $H_1$ in the range of 
$ (13 - 27)~ R_*$.
This size agrees with that determined for the Alfv\'en radius, i.e.
 in the range $ (14 - 30)~ R_*$.
By  interpreting   the rapid decay of the
flare  
as leakage of the emitting electrons from the   
trapping magnetic structure of size $H_1$,  we determine  $H_1$ in the range ($10-20)~ R_*$.
%
As a result, one of the two stars of the system might have a corona of a few
stellar radii only, and  interbinary collisions can still occur because
of a large (i.e. average size $H_1 \sim 19 R_*$)
helmet streamer located  on the other  star.

The leakage model
has  shed light 
on the nature of the electrons responsible for
the flare.
We state  that interbinary   collisions
are able to accelerate electrons to
relativistic velocities. In fact, we determine
a Lorentz factor  $\gamma$ up to 632.
This confirms  that the emission in this pre-main sequence system
is   synchrotron radiation, as  postulated from
the  observed linear polarization (Phillips et al. 1996).

In particular, streaming  back and forth between the two mirror points
(P$_0$ and P$_1$ in Fig. 4) of the helmet streamer,
with magnetic fields between
0.12 G and 125 G,
the relativistic electrons  produce
the observed synchrotron radiation.
The trapping by the helmet streamer is, however, only partial
and during each reflection
particles escape and the emission fades out relatively quickly.
Leakage  explains
the puzzling finding that flares have similar decay times
(1 -- 2 hours), not only in different radio bands, but also in the millimeter band.
The latter similarity would be difficult to
explain on the basis of any loss mechanism dependent on the electron energy.

\begin{acknowledgements}

We thank  the referee 
for reading the manuscript
with great attention and
providing many constructive comments and critical remarks.
We acknowledge the IRAM staff from the Plateau de Bure and from Grenoble,
in particular Jan Martin Winters,
for their help provided during the observations and data reduction.
IRAM is an international institute for research in millimeter
astronomy funded by the
Max Planck Gesellschaft, Germany,
the Centre National de la Recherche Scientifique, France,
and the Instituto Geografico Nacional, Spain.
\end{acknowledgements}

\appendix
\section{energetic considerations}

Accelerated non-thermal electrons are
subject to energy losses both because synchrotron radiation is emitted and
because of collisions with thermal electrons.
In addition, energy losses are also due to the
interaction of non-thermal electrons by inverse Compton (IC)
processes with the stellar radiation field (external IC) and even
with  photons emitted by the synchrotron process itself in the radio band
(synchrotron self-Compton (SSC)).

If synchrotron
losses are responsible for the  observed decay  ($\tau$, in hours),
the following relationship holds between magnetic field intensity $B$ (in Gauss)
and Lorentz factor $\gamma$
(Blumenthal \& Gould 1970):
\beq
\gamma B^2={2.2 \times 10^5 \over \tau_{\rm synchrotron}}.
\label{b_ga2}
\eneq
%
From  Eqs. (\ref{b_ga2})  and (\ref{b_ga}),
with $\tau_{\rm synchrotron}=2.31$ hours, the value
of the magnetic field strength
in the emitting region and the
value of the Lorentz factor of the emitting electrons can be derived
as:
$B=56$~G and $\gamma$=30.

A magnetic dipole field, with $B(R_*)$ at the stellar surface,
attains the intensity of 56 G
at a distance $H/R_*=[B(R_*)/56]^{1\over 3}$. 
In the literature, there are very few estimates of
the magnetic flux $B(R_*)$ in   weak T Tauri stars.
Basri et al. (1992) determined the upper limit of 1500 G
for the product B${\it f}$, where ${\it f} <$1, for  TAP 35. An even higher value (B${\it f} >$ 2000 G) is derived by
Guenther et al. (1999) in the source LkCa 16. 
Basri et al. (1992) exclude fields above 3000 G, and 
more recently (Bower et al. 2003), the value of B=2600 G has been measured 
in a weak T Tauri in the Orion Nebula. Therefore,  1000 -- 3000 G seems to be a likely range.
For $B(R_*) \leq$ 3000 G,
the magnetic dipole field attains the intensity of 56 G
at a distance $H/R_* \leq 3.8$ from the stellar surface.
This scenario cannot account for the observed extended radio emission (\S 2).
We also exclude  synchrotron losses as being responsible for
the lifetime  $\tau=1$ hour of the Phillips et al. (1996) observations
at
$\nu_0=8.2$ GHz.
The application of Eqs. (\ref{peak}) and (\ref{b_ga2}) would
give $B \sim 223$ G and $\gamma =4$, and  hence
emission  coming from a region relatively close to the stellar surface
($H/\rm R_* \leq 2.4 $)
and from mildly relativistic electrons, in contrast with
previously  reported considerations (\S 1 and 2).

IC losses  cannot account for the observed emission
decay either. In fact, as derived hereafter,
IC losses are
even lower than synchrotron losses because
the radiation energy density ($U$) is smaller than
the magnetic energy density
i.e.,
\beq
U< B^2/8\pi.
\label{compton}
\eneq
A dipole magnetic field, with the intensity
at the surface of the star in the range quoted above
(say 2 10$^3$ G), has
$B^2/8\pi\simeq 
1.6 \times 10^5 (H/R_*)^{-6}$ erg cm$^{-3}$,
while  $U=L_*/4\pi c H^2
=0.86~(H/R_*)^{-2}$
erg cm$^{-3}$ for known values
of radius and luminosity of V773 Tau A ($R_* \sim 2.4 R_{\sun}$,
$L_* \sim 3 L_{\sun}$) (Skinner et al. 1997; Welty 1995).
Hence, if the radiation of the star
is the source of the seed photons (IC), condition (\ref{compton}) holds
for $H \leq 21 R_*$.
On the other hand, for  SSC,
the related flare luminosity is $L_{\rm flare}=6.4 \times 10^{26}$ erg~sec$^{-1}$,
which,
supposing that the flare takes place in a spherical region
of radius $a R_*$, with $a >0.01$, gives  $U= 6~10^{-8} a^{-2}$ erg cm$^{-3}$.
Hence, condition (\ref{compton})
holds  throughout the region of interest,
since  $(H/R_*) \leq 118~a^{1/3}$ is verified for every 
$(H/R_*) \leq 25$.

Finally, the thermalization time due to collisions of relativistic electrons
in an ionized gas of density $n$ is given by
(Petrosian 1985; Massi \& Chiuderi-Drago 1992):
\beq
\tau_{\rm collisions}\simeq 4.16 \times  10^{8} {\gamma\over n},
\label{coll}
\eneq
where $n$ is  in cm$^{-3}$ and $\tau$ is expressed in hours.
Therefore, collisions would be important
only  for $n \sim  \gamma~1.8 \times 10^8$ cm$^{-3}$,
 which,
for the relativistic electrons ($\gamma >>$1) implied by the linear polarization,
 does not  match the condition derived from Faraday depolarization
($n < ~1. \times 10^9$ cm$^{-3}$; see \S2).

\end{document}